\documentclass[12pt]{iopart}
\usepackage{graphicx}

\begin{document}

\title{Non-centro-symmetric superconductors $Li_2Pd_3B$ and $Li_2(Pd_{0.8}Pt_{0.2})_3B$: amplitude and phase fluctuations analysis of the experimental magnetization data}

\author{ P. Badica$^1$, S. Salem-Sugui, Jr.$^2$, A.D. Alvarenga$^3$, and G. Jakob$^4$}

\address{$^1$National Institute of Materials Physics, Bucharest-Magurele, Atomistilor 105bis, 077125 Romania}
\address{$^2$Instituto de Fisica, Universidade Federal do Rio de Janeiro,
21941-972 Rio de Janeiro, RJ, Brazil}
\address{$^3$Instituto Nacional de Metrologia Normaliza\c{c}\~ao e
Qualidade Industrial, 25250-020 Duque de Caxias, 
RJ, Brazil}
\address{$^4$Institute of Physics, Mainz University, Staudingerweg 7, 55127 Mainz, Germany}

\begin{abstract}
We report on magnetisation data obtained as a function of temperature and magnetic field in $Li_2(Pd_{0.8}Pt_{0.2})_3B$ and $Li_2Pd_3B$. Reversible magnetisation curves were plotted as $\sqrt{M}$vs. T. This allow study of the asymptotic behavior of the averaged order parameter amplitude (gap) near the superconducting transition. Results of the analysis show, as expected, a mean field superconducting transition for $Li_2Pd_3B$. On contrary, a large deviation from
the mean field behavior is revealed for $Li_2(Pd_{0.8}Pt_{0.2})_3B$. This is interpreted as due to
the strength of the non s-wave spin-triplet pairing in this Pt-containing compound which produces nodes in the
order parameter and consequently, phase fluctuations. The diamagnetic signal above $T_c(H)$ in $Li_2Pd_3B$ is well explained by superconducting Gaussian fluctuations, which agrees with the observed mean field transition. For $Li_2(Pd_{0.8}Pt_{0.2})_3B$ the diamagnetic signal above $T_c(H)$ is much higher than the expected Gaussian values and appears to be well explained by three dimensional critical fluctuations of the lowes-Landau-level type, which somewhat agrees with the scenariou of a phase mediated transition. 

\end{abstract}
\maketitle

\section{Introduction}
Superconductivity was found in $Li_2(Pd_{1-x}Pt_{x})_3B$ x=0-1 system \cite{badica1,badica2,badica3}. This sytem shares properties with low-$T_c$ superconductors due to its low-$T_c$ and with high-$T_c$ due to its perovskite-like structure. It has a non-centro-symmetric lattice structure  producing an asymmetric spin-orbit-coupling which violates Pauli's parity principle  allowing admixture of spin-singlet and spin-triplet states in the superconductor order parameter. \cite{spin1,spin2,spin3,hai-hu,samokhin} Such admixture of states with different symmetries induces an anisotropy in the order parameter, which depending on the strength of the asymmetric spin-orbit-coupling can lead to the existence of line nodes.\cite{spin3}   As it is well know, the existence of line nodes in the order parameter allows quasi-particles excitations at low temperatures, affecting the density of states. \cite{spin3} Line nodes also play an important role on the order parameter fluctuations near $T_c$, since phase and amplitude fluctuations have different contributions in the node and in the anti-node producing a change in the density of states near $T_c$. \cite{kwon} Such a change in the density of states changes the shape of the asymptotic behavior of the order parameter near $T_c(H)$ producing a deviation from the expected mean field behavior $\approx (T-T_c)^{1/2}$. \cite{kwon,ana,jesus} 

In  this work we study the asymptotic behavior of the order parameter in $Li_2Pd_3B$ and $Li_2(Pd_{0.8}Pt_{0.2})_3B$ near $T_c(H)$ using precision magnetisation measurements obtained as a function of temperature and magnetic field. The work addresses the existence of phase fluctuations in the vicinity of $T_c$ which are expected if the gap contains line nodes due to the spin-triplet state admixture in the studied systems. We also analyse the diamagnetic magnetisation found near and above the transition temperature, $T_c(H)$, on both samples  by fitting the diamagnetic signal above $T_c(H)$ with the three-dimensional Gaussian fluctuation Schimidt expression, which explains most of the diamagnetic fluctuations above $T_c$ in low-Tc (S-wave BCS) superconductors)\cite{tinkham},  and by means of a lowest-Landau-level scaling analysis \cite{welp,rosenstein,tesa} (critical fluctuations) performed on data near $T_c(H)$.  
The work is motivated by reported deviations from the BCS-like expected behavior observed at low temperatures for the penetration depth \cite{spin1} and for the Knight shift \cite{spin2} on $Li_2Pt_3B$ which strongly suggested the existence of line nodes in this compound. Also, it is interesting to investigate if the strength of the asymmetric spin-orbit coupling (or the strength of the spin-triplet admixture) observed in $Li_2Pt_3B$ holds for $Li_2(Pd_{0.8}Pt_{0.2})_3B$. To the knowledge of the authors, there is no report in the literature devoted to diamagnetic fluctuations studies in the $Li_2(Pd-Pt)_3B$ system. 
\section{Experimental}
The experiment was conducted by careful and precise measurement of magnetization data by a 5T MPMS Quantum-Design magnetometer on $Li_2Pd_3B$ ($T_c$=8.1 K) and  $Li_2(Pd_{0.8}Pt_{0.2})_3B$ ($T_c$=5.8 K) samples. Samples were obtained by arc melting as described in Refs.\cite{badica1,badica2,badica3} and are of high-quality each one exhibiting a single phase and sharp transition.  All magnetisation data were obtained with sample cooled from temperatures well above $T_c$ in zero applied magnetic field. Magnetic field was applied in the no overshoot mode and data was obtained by heating the sample with fixed increments of temperature ( 0.1 K in the region near $T_c(H)$). We also obtained field-cooled magnetisation curves, M vs. T, in order to obtain the reversible magnetisation. Data were obtained from 1.8 K up to temperatures well above $T_c(H)$ allowing to determine the normal state-magnetisation, Mback, which was subtracted for each correspondent curve. 

\section{Results and discussion}
\begin{figure}
\includegraphics[width=\linewidth]{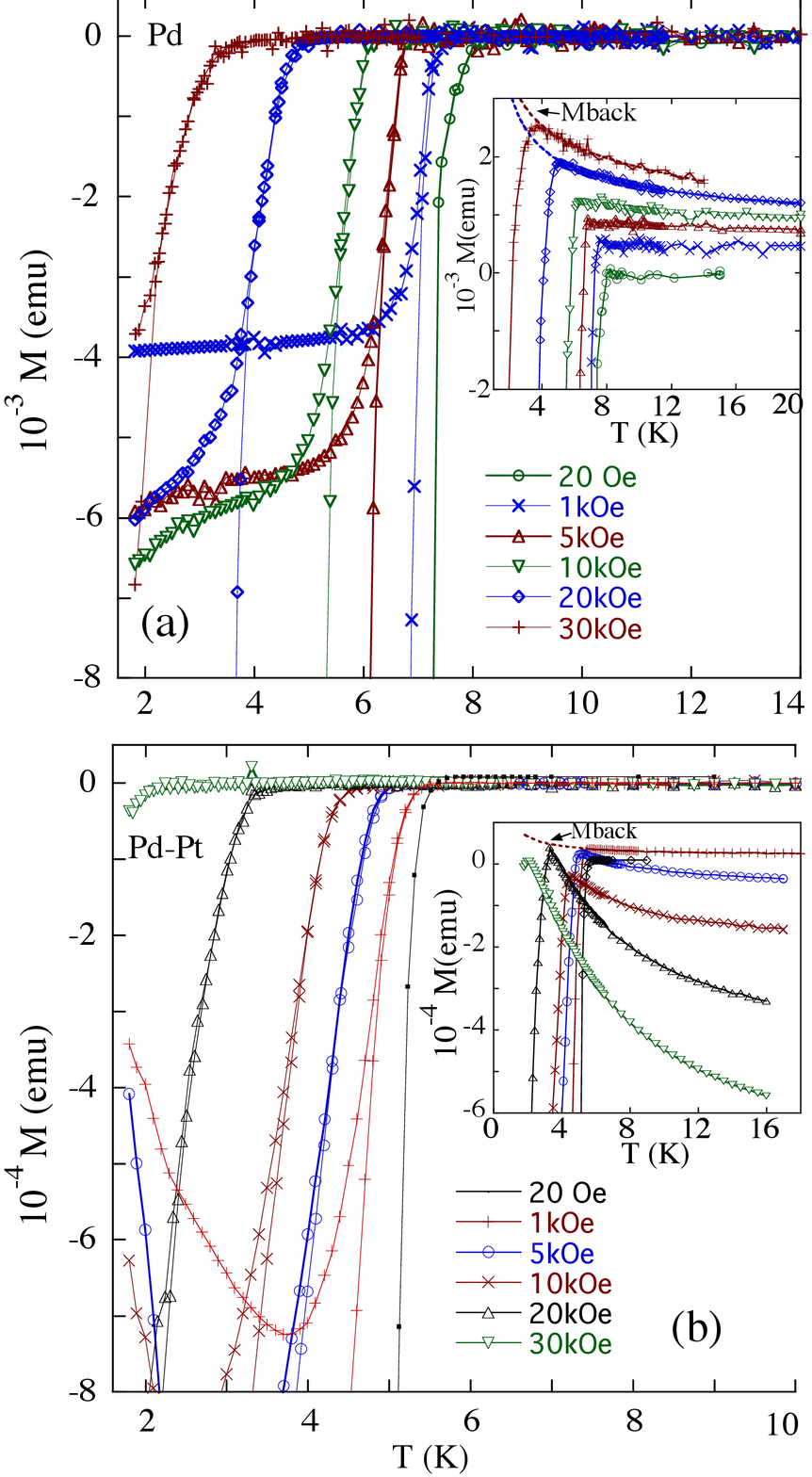}
\caption{Isofield magnetisation curves for: a)$Li_2Pd_3B$ and b)$Li_2(Pd_{0.8}Pt_{0.2})_3B$. The insets show magnetisation curves as obtained prior to background subtraction. }
\label{fig1} 
\end{figure}
 
 Figure 1a and 1b show magnetisation curves for both samples obtained after the proper background correction. The insets of Fig.1a and 1b show data prior to background correction. We observe that the normal state magnetisation for both samples follows a Curie-Weiss type of the form, $Mback=a+b/T$, but only for magnetic fields below 20 kOe. Above 20 kOe, an additional linear term, cT, had to be include for $Li_2Pd_3B$ (small correction). For $Li_2(Pd_{0.8}Pt_{0.2})_3B$ for fields 20 kOe (very small correction) and  30 kOe (small correction) we used $Mback=a+(b+clog(T))/T$, where the additional (log(T))/T term was included to account for the sample holder signal \cite{holder}. This expression for $Mback$ fitted quite well the normal-state region curves. To check for this backgound correction we measured an isothermal hysteresis curve, M vs. H, in $Li_2(Pd_{0.8}Pt_{0.2})_3B$ at 3.5 K, which shows magnetisation values that reasonably agree with the correspondent values in the corrected $Mvs.T$ curve taken at H=30 kOe.  An interesting finding about the normal state magnetisation on both samples as shown in the insets of Figs. 1a and 1b, is that while the value of the constant $a$ is positive (paramagnetic) and increases with field for $Li_2Pd_3B$, it is negative (diamagnetic) for $Li_2(Pd_{0.8}Pt_{0.2})_3B$ with diamagnetism increasing with field, as commonly found in high-$T_c$ superconductors. \cite{rosenstein2}

In the next we shall analyze the reversible magnetization. Since we are interested to study possible effects of phase fluctuation near $T_c(H)$ it is more conveniente to plot the obtained magnetisation curves as $\sqrt{M}$ vs. T, this is because near $T_c(H)$ the quantity $\sqrt{M}$ is directly proportional to the superconducting order parameter $\psi$ by the relation \cite{deGennes}
\begin{equation} 
M=-\frac{e \hbar}{mc}|\psi|^2 .
\end{equation}
The well known equation (1) above was obtained upon application of the Abrikosov approximation \cite{abrikosov} to the Ginzburg-Landau equation. 
Curves of $\sqrt{M}$ vs. T near $T_c(H)$ allow to study the asymptotic behavior of the order parameter which can be expressed as $\sqrt{M} \propto [T_c(H)-T]^m$ where $T_c(H)$ is the mean field transition temperature and $m$ is the mean field exponent. The theoretical mean field exponent value is m=1/2,  for both $s$-wave BCS superconductors \cite{deGennes}, and $d$-wave superconductors within a Ginzburg-Landau theory \cite{xu}. Since the studied system here have a low $T_c$ one would expect that neither amplitude nor phase fluctuations should play an important role, and the resulting transition should be of the mean-field type \cite{emery}, with $m\approx 1/2$. However, the existence of line nodes in the order parameter can turn phase fluctuations important. It has been shown in ref.\cite{kwon} that the presence of nodes and anti-nodes in the order parameter has definite consequences for the phase and amplitude fluctuations, which can have an effect on the superfluid density of states, reducing the gap in the vicinity of $T_c$. Such change in the gap may alter the expected mean field value of the exponent $m$, generating a larger value. \cite{ana,jesus}
In Figs. 2 and 3 are shown the resulting $\sqrt{M}$ vs. T curves using the data from Figs. 1a and 1b respectively. For each curve we separate the region below $T_c(H)$ where phase fluctuations may play an important role from the region above $T_c(H)$ where amplitude fluctuations generate an  anomalous enhancement of the magnetisation. The analysis is focused only in the reversible regime, and for this reason only $M vs.T$ curves obtained with fields above 5 kOe are analyzed. Values of $T_c(H)$ are estimated in each curve from the extrapolation of the linear magnetisation to zero (Abrikosov method \cite{deGennes}).

\begin{figure}[t]
\includegraphics[width=\linewidth]{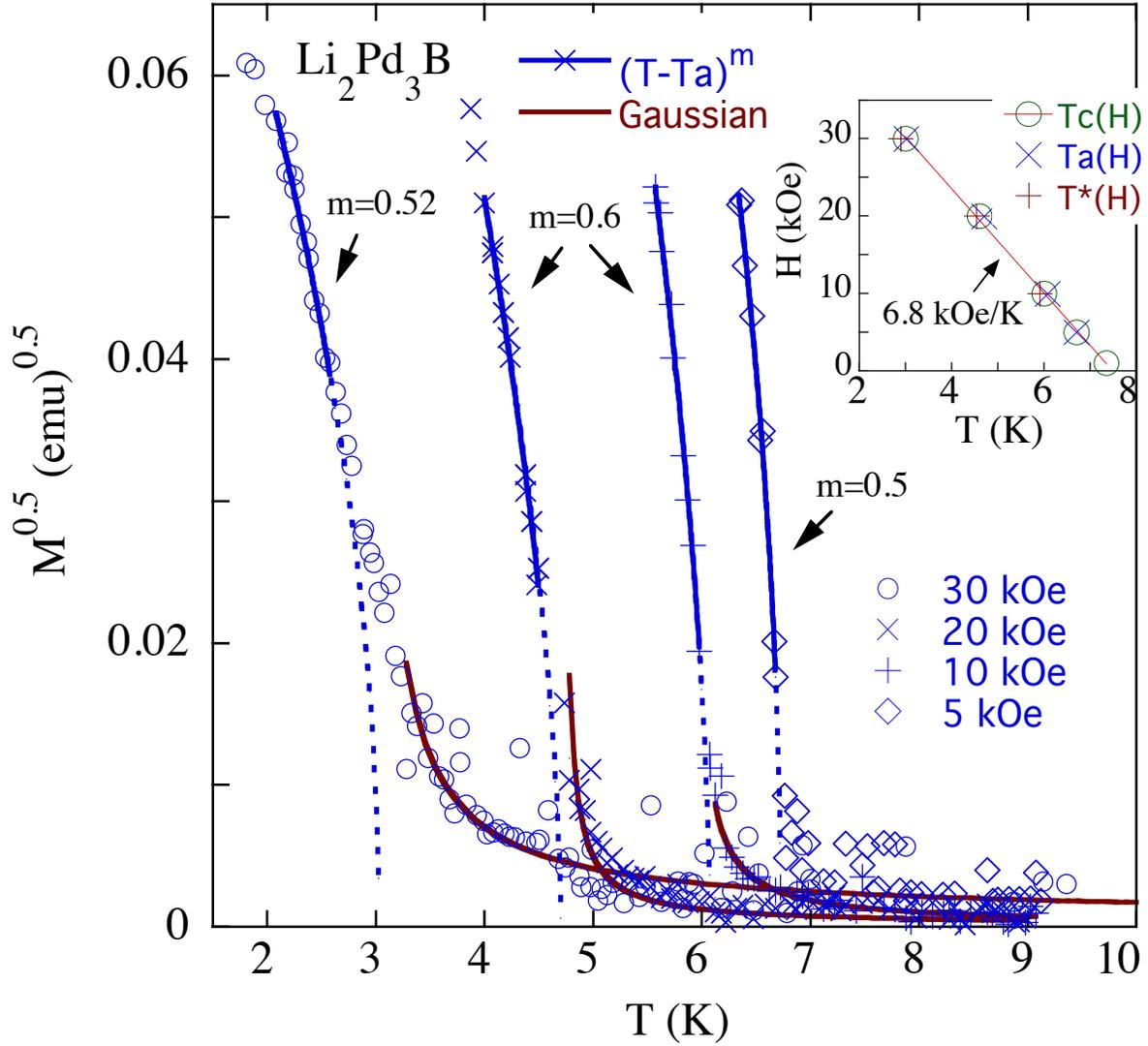}
\caption{Isofield curves of $\sqrt{M}$ vs. $T$ for $Li_2Pd_3B$. The inset show $T_a(H)$, $T_c(H)$ and $T^*(H)$ plotted against H.}
\label{fig2}
\end{figure}

\begin{figure}[t]
\includegraphics[width=\linewidth]{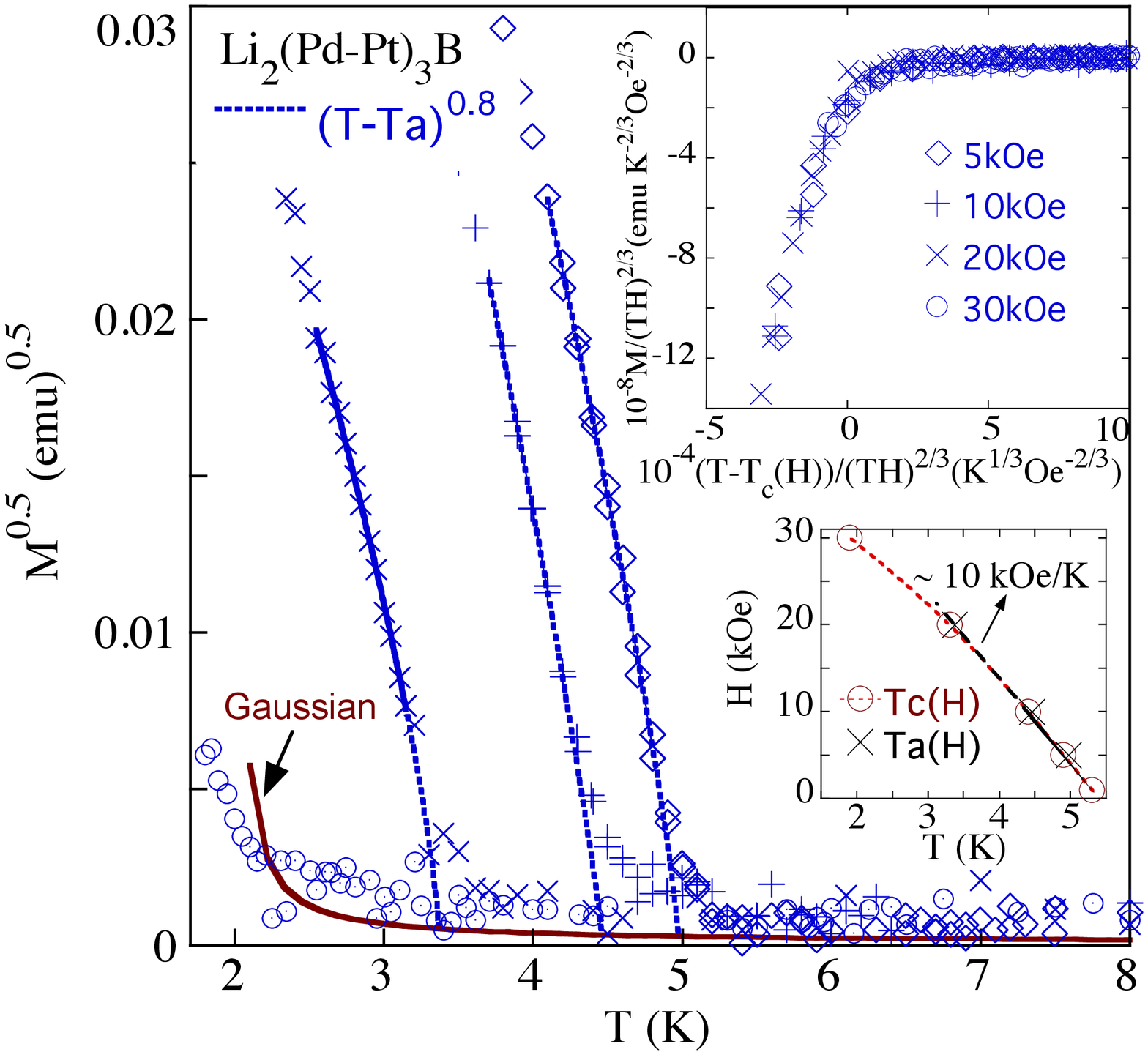}
\caption{Isofield curves of $\sqrt{M}$ vs. $T$ for $Li_2(Pd_{0.8}Pt_{0.2})_3B$. The upper inset show the lowest-Landau-level analysis results. The lower inset show $T_a(H)$ and $T_c(H)$ plotted against H.}
\label{fig3}
\end{figure}

We first discuss the general scaling analysis used to fit the region below $T_c(H)$. This region for each curve is delimited from below by a temperature below which the Abrikosov approximation can not be applied, and from above by a change in the curvature of the curve, or inflection point, occurring near $T_c(H)$. 
Within this region, each curve as appearing in Figs. 2 and 3 was fittted to the general form $\sqrt{M} \propto [T_a(H)-T]^m$ where $T_a(H)$ is an apparent transition temperature, and $m$ is a fitting exponent. Deviations of the exponent $m$ from the mean field value 1/2 may infer a phase mediated transition. 
Results of the fittings are shown for each curve. Resulting values of $T_a(H)$ for both samples are virtually the same as obtained from the Abrikosov method mentioned above. So this method applied to the studied samples produces consistent values of the transition temperature. The consistence of the $T_a(H)$ values is shown in the insets of Figs. 2 and in the lower inset of fig. 3 where values of $T_a(H)$ are plotted with values of $T_c(H)$. Resulting values of the exponent $m$ for $Li_2Pd_3B$ suggests a typical mean field behavior  (despite two values of $m$ are slightly larger than 1/2, these values are well within the expected mean field value found for instance in Nb \cite{ana}). But, resulting values of $m$ for $Li_2(Pd_{0.8}Pt_{0.2})_3B$ are much larger than 1/2 ($m \approx 0.8$ for all curves), and as discussed above, suggest a phase mediated transition, which is consistent with the existence of nodes in the order parameter. These results suggest that the strength of the admixture of singlet and triplet spin states in the order parameter is not negligible for $Li_2(Pd_{0.8}Pt_{0.2})_3B$ but it is for $Li_2Pd_3B$. Probably the admixture of triplet to singlet states in $Li_2Pd_3B$ produces some anisotropy in the order parameter, but our analysis support the idea that there are no nodes.

We now discuss the anomalous enhancement of magnetisation occuring in the vicinity and above $T_c(H)$ as observed in Figs. 2 and 3. 
Our first attempt was to fit the diamagnetic signal above $T_c(H)$ appearing on the curves of Figs. 2 and 3 with the three dimensional Gaussian-fluctuation expression derived by Schimidt \cite{tinkham} 
$$M_{flu}\approx T/(T-T^*)^2$$
$T^*$ is a fitting parameter and it is expected to coincide with $T_c(H)$. The Gaussian-type of diamagnetic fluctuations above $T_c(H)$ is commonly observed in many superconductors systems of low-$T_c$  \cite{tinkham,jesusPbIn} and also of some high-$T_c$ (after taking off the short-wavelength fluctuations)  \cite{jesusLa}. Results of the fittings with  Gaussian formula (2) for the magnetisation fluctuation on $Li_2Pd_3B$ curves are shown in Fig. 2. The fittings are of excellent quality and the estimated values of $T^*$ are virtually equal to the values of 
$T_c(H)$. $T^*(H)$, $T_a(H)$ and $T_c(H)$ are presented in the inset of Fig. 2. On the other hand, Gaussian expression (2) failed to fit the fluctuation magnetisation above $T_c(H)$ on the curves of $Li_2(Pd_{0.8}Pt_{0.2})_3B$. The fittings conducted on the curves of Fig. 3 for data in the region above $T_c(H)$ produced low-quality fittings with values of $T^*$ much smaller than the correspondent values of $T_c(H)$. For example, we present a fit conducted on the 30 kOe data above 2K where we fixed the value of $T^*$= 2 K. It is possible to see from this fitting that the fluctuation magnetisation observed above $T_c(H)$ for $Li_2(Pd_{0.8}Pt_{0.2})_3B$ is much larger than the contribution due solely to the Gaussian-type of diamagnetic fluctuation. Larger diamagnetic fluctuations below and above $T_c(H)$ for high magnetic fields have been observed in many superconductor layered systems with large values of the Ginzburg-Landau parameter $\kappa$ \cite{klemm,rosenstein2}. This situation is commonly explained in terms of lowest-Landau-level, (LLL),  fluctuations theories. LLL theories consider fluctuations-fluctuations interactions within the Ginzburg-Landau formalism and predict specific scaling-laws that should be obeyed by the magnetisation and temperature \cite{rosenstein2}. To perform the scaling \cite{welp,rosenstein} one should replace the 
temperature $x-axis$ and magnetisation $y-axis$ of each curve to the respective scaling forms $(T-T_c(H))/(TH)^{3/2}$ and $M/(TH)^{3/2}$ and plot together all scaled  curves. This scaling is appropriate to three-dimensional systems, as is the present case. The only free parameter in the scaling procedure is the mean field temperature $T_c(H)$ which is adjusted for each curve so that all results fall onto a single universal curve. We applied this scaling approach on the reversible data of $Li_2(Pd_{0.8}Pt_{0.2})_3B$ (Fig. 1b for fields of 5, 10, 20 and 30 kOe). The result is presented in the upper inset of Fig. 3. One may observe that the collapse of the different M vs. T curves is almost perfect. Values of $T_c(H)$ obtained from the lowest-Landau-level scaling analysis virtually coincides 
with values of $T_a(H)$ (lower inset of Fig. 3b). This result evidences the importance of lowest-Landau-level fluctuations on $Li_2(Pd_{0.8}Pt_{0.2})_3B$. Importantly, the enhanced diamagnetic signal observed above $T_c(H)$ in $Li_2(Pd_{0.8}Pt_{0.2})_3B$, which values are much higher than that expected from Gaussian-fluctuations, is very likely to be related to the strength of the spin-triplet admixture in to the order-parameter, due to Pt, which enhances the asymmetric spin-orbit-coupling when added (by substitution) to $Li_2Pd_3B$ system.  It should be noted that a similar scenario occurs in high-$T_c$ superconductors where phase fluctuations near $T_c(H)$ is accompanied by enhanced amplitude fluctuations  near and above $T_c(H)$. \cite{beck,ana}  We also mention that we applied the same LLL scaling approach on the reversible data of $Li_2Pd_3B$ (not shown) producing a poor collapsing of the curves, which somehow agrees with the observation that the fluctuation magnetisation above $T_c(H)$ in this system is well explained in terms of Gaussian fluctuations, as addressed above.

\section{Conclusions}
Our work presents phase and fluctuation analysis results deduced from the experimental reversible magnetisation data measured on the $Li-Pd-Pt-B$, x=0 and x=0.2, non centro symmetric superconducting system.Results point on the existence of line nodes in the order parameter of the $Li_2(Pd_{0.8}Pt_{0.2})_3B$ and a pure s-wave for the $Li_2Pd_3B$ sample as for Nb, confirming similar conclusions obtained from penetration depth, specific heat and Knight shift measurements \cite(spin1,spin2). The study also support the idea that the strength of the spin-triplet admixture in the order parameter of $Li_2(Pd_{0.8}Pt_{0.2})_3B$ plays an important role on the amplitude and phase fluctuations near $T_c(H)$. Diamagnetic fluctuations observed above $T_c(H)$ in $Li_2Pd_3B$ can be well explained by Gaussian fluctuations, while for $Li_2(Pd_{0.8}Pt_{0.2})_3B$ the diamagnetic signal above $T_c(H)$ for higher fields are much higher than the values expected by Gaussian fluctuations and likely well explained by critical three-dimensional lowest-Landau-level fluctuations. Amplitude and phase fluctuation scaling analysis of the experimental magnetization data turns to be a convenient and relatively simple method to study unconventional systems as in this case of the non centro symmetric superconductors showing particular non s-wave features.

PB acknowledges support from Alexander von Humboldt Foundation and CNCSIS-UEFISCSU (PNII PCE 513/2009 and PNII PCCE 239/2008-9/2010). SSS and ADA acknowledge support from CNPq.


\begin{thebibliography}{10}
\bibitem{holder}A (logT)/T term in the magnetization was observed when measuring the sample holder used for sample containing Pt.
\end{thebibliography}
\end{document}